# Stabilization of Ferroelectric Domains via Interfacial Strain in $HoFeO_3$- $BaTiO_3$ Nanofiber Heterostructures for Enhanced Hybrid Energy Harvesting

Smita Chaturvedi*[1,2], Swathi Krishna[2], Shivshankar Jokare[3], Supriya Sahoo[4], Vikash Kushwaha[4], Priyank Shyam[5], Mandar Shirolkar[6], and R. Boomishankar[4]

*[1] Nayanta University, Pune 411021, India.*
*[2] Department of Physics, Indian Institute of Science Education and Research, Pune 411008, India.*
*[3] Department of Physics, Savitribai Phule Pune University, Pune 411007, India.*
*[4] Department of Chemistry, Indian Institute of Science Education and Research, Pune 411008, India.*
*[5] Arhus University, Denmark*
*[6] Norel Nutrient Bio-Agro Tech Pvt. Ltd, Baner, Pune 411045, India.*

***Correspondence:** Smita Chaturvedi (*smita.chaturvedi24@gmail.com*)

## Abstract

Piezoelectric nanogenerators (PENGs) are an attractive platform for self-powered mechanical-energy harvesting, but their output is often limited by depolarization fields, charge leakage, and inefficient stress transfer to the active dipoles, particularly in polymer-dominated composites. Here we report $HoFeO_3$-$BaTiO_3$ (HFO-BTO) nanofiber heterostructures prepared by alternating sol-gel electrospinning, in which orthorhombic $HoFeO_3$ (*Pbnm*) and tetragonal $BaTiO_3$ (*P4mm*) coexist in intimate contact. Rietveld-refined X-ray diffraction, high-resolution transmission electron microscopy, and Raman spectroscopy indicate interfacial lattice distortion. Polarization–electric field loops show that the heterostructure retains a polarization comparable to that of $BaTiO_3$ (16.39 vs 15.80 $\mu C\ cm^{-2}$ at zero field) while its switching-current maximum moves to a higher field (~7 vs ~3.5 $kV\ cm^{-1}$), consistent with strain-stabilized ferroelectric domains. When the nanofibers are encapsulated in polydimethylsiloxane (PDMS) and excited under cyclic impact, the device produces a hybrid output: the piezoelectric response of the strained $BaTiO_3$ is supplemented by a triboelectric contribution from contact electrification. The HFO-BTO device delivers open-circuit voltage peaks of about +6 V and −8 V and a power density of 24.31 $\mu W\ cm^{-3}$ at a 4.7 MΩ load. This is 2.4 times the value for an identically fabricated $BaTiO_3$ device (10.00 $\mu W\ cm^{-3}$) and 8.7 times that of the $HoFeO_3$ device (2.78 $\mu W\ cm^{-3}$). These results link interface-driven lattice distortion to improved electromechanical output and identify all-oxide nanofiber heterostructures as a promising route to lead-free energy harvesting.

## Introduction

The rapid proliferation of self-powered sensors, wearable electronics, structural-health-monitoring systems, and Internet-of-Things (IoT) nodes compounded by the explosive growth of artificial-intelligence and edge-computing infrastructure has sharply increased the demand for compact, autonomous, and maintenance-free power sources for distributed electronics. [1-3] Conventional electrochemical batteries suffer from finite lifetime, periodic replacement, and the environmental burden associated with toxic waste and raw-material extraction, making them poorly suited to self-sustained distributed systems. Harvesting ambient mechanical energy offers an attractive route to perpetual, battery-free operation. [4]

Among mechanical-to-electrical conversion strategies, piezoelectric nanogenerators (PENGs) are especially appealing because they convert ubiquitous mechanical stimuli vibration, impact, and human body motion directly into electricity through strain-induced polarization change, requiring neither an external bias nor complex conversion circuitry. [1,4] Since Wang and Song first demonstrated nanoscale piezoelectric power generation from ZnO nanowires in 2006, [1] lead-free perovskite ferroelectrics, particularly $BaTiO_3$, have become benchmark PENG materials owing to their large piezoelectric coefficient ($d33$ up to ~190 pC $N^{-1}$ in bulk ceramics), high dielectric constant, and excellent chemical stability. [5-8]

$BaTiO_3$ adopts the $ABO_3$ perovskite framework, in which $Ba^{2+}$ occupies the 12-coordinate A-site and $Ti^{4+}$ resides within a corner-sharing $TiO_6$ octahedron. Below the cubic-to-tetragonal Curie temperature (~120 °C), the $Ti^{4+}$ ion spontaneously displaces off-center along the crystallographic [001] axis by approximately 12 pm, stabilizing the non-centrosymmetric *P4mm* phase and giving rise to a net spontaneous polarization of ~26 μC $cm^{-2}$ in single crystals. [5,9] First-principles density functional theory (DFT) studies by Cohen [9] demonstrated that this ferroelectric instability originates predominantly from Ti 3*d*-O 2*p* covalent hybridization rather than from purely ionic electrostatic forces: the hybridization lowers the energy of the off-center configuration, and any perturbation that strengthens the Ti-O orbital overlap (such as compressive epitaxial strain or interfacial charge redistribution) directly amplifies the displacement and the resulting polarization. [9-11] Being fully lead-free, $BaTiO_3$ also constitutes an environmentally benign alternative to the widely used but toxic $Pb(Zr,Ti)O_3$ (PZT) system. [5]

Despite this promise, the practical performance of oxide-based PENGs remains constrained by several intrinsic loss channels. A substantial fraction of the mechanically induced polarization is rendered electrically inactive by: (i) *parasitic (non-switchable) polarization* from pinned domains and charge-trapped defects; (ii) internal *depolarization fields* that oppose the spontaneous polarization in thin or nanostructured volumes; (iii) *charge leakage* through grain-boundary conduction pathways; and (iv) *inefficient stress transfer* to the active dipolar regions due to mechanical compliance mismatches. [6,12,13] Collectively, these effects depress the achievable voltage and power density and hinder real-world deployment of oxide-based harvesters.

Nanoscale heterostructure has emerged as a powerful strategy to mitigate these losses. Multilayer, superlattice, and strain-engineered perovskite systems have demonstrated that controlled interfacial lattice distortion can simultaneously amplify the useful switchable polarization and suppress leakage currents by modifying the local bonding environment and electronic structure of the active layer. [11,14,15] The central design principle is interface engineering by coupling two structurally dissimilar perovskites or perovskite-related oxides across a coherent or semi-coherent interface, epitaxial strain modifies the local bond lengths and angles in a way that cannot be achieved by bulk compositional tuning alone.

Rare-earth orthoferrites ($RFeO_3$, R = rare earth) and $BaTiO_3$ form a structurally and functionally compelling pairing for interface-engineered PENGs. Orthorhombic $HoFeO_3$ provides mechanically rigid $FeO_6$ octahedra and is a canted G-type antiferromagnet below its Néel temperature (~640 K), and its corner-sharing octahedral framework is thermally and chemically robust at the processing temperatures used here. [16-19] Tetragonal $BaTiO_3$ provides the primary ferroelectric functionality through off-center $Ti^{4+}$ displacement within $TiO_6$ octahedra. [5,9] When these two phases are coupled in a heterostructure nanofiber architecture, interfacial strain modifies the local bonding and lattice symmetry of both constituents, enabling enhanced polarization utilization while mitigating parasitic-polarization losses. Orthoferrite-ferroelectric fiber heterostructures have also been realized by electrospinning, for example $NdFeO_3$-$PbZr_{0.52}Ti_{0.48}O_3$ core-shell nanofibers designed for strain-mediated magnetoelectric coupling. [20]

Morphology is equally decisive for PENG performance. One-dimensional anisotropic architectures nanofibers and nanowires offer clear advantages over thin films and particulate assemblies, including superior uniaxial stress alignment, a large surface-to-volume ratio, and highly efficient mechanical load transfer along the fiber axis. [21,22] Electrospinning provides a scalable, cost-effective route to continuous, high-aspect-ratio oxide nanofibers with intimate phase integration and built-in geometric anisotropy; the sol-gel precursor chemistry is flexible enough to accommodate multi-component compositions such as the present HFO-BTO system. [20,23,24] It is important to note that at the nanoscale, ferroelectric distortions are frequently local in character: long-range X-ray diffraction may report an apparently pseudo-cubic or disordered pattern even when short-range Ti off-centering persists robustly at the unit-cell level, so that finite-size effects alone need not extinguish the underlying tetragonal spontaneous polarization. [25]

When a piezoelectric nanofiber mat is encapsulated in a soft elastomeric matrix (PDMS) and operated in contact-separation mode, contact electrification at the fiber-polymer and film-electrode interfaces generates an additional triboelectric charge that couples additively with the piezoelectric response. Incorporating a high-permittivity ferroelectric oxide into the PDMS matrix simultaneously raises the piezoelectric potential and the effective dielectric permittivity of the composite (Maxwell-Wagner-Sillars interfacial polarization), reinforcing the triboelectric charge storage capacity; such piezo-triboelectric hybridization has been repeatedly shown to boost the net output of $BaTiO_3$/PDMS generators significantly beyond what either mechanism alone can achieve. [26-33]

In this work, we prepare $HoFeO_3$-$BaTiO_3$ (HFO-BTO) nanofiber heterostructures by alternating sol-gel electrospinning and evaluate them as the active phase of PDMS-encapsulated nanogenerators. The aim is to carry the interface-engineering principles established for epitaxial oxide films over to a one-dimensional nanofiber architecture. In this architecture, interfacial strain between the two oxides tunes the local $TiO_6$ environment, while the fiber geometry aids stress transfer. Using structural, spectroscopic, first-principles, magnetic, and electrical measurements, we show that the heterostructure device delivers a power density of 24.31 $\mu W\ cm^{-3}$ at a 4.7 MΩ load, 2.4 times that of an identically fabricated $BaTiO_3$ device. We then analyze how interfacial distortion, strain-stabilized ferroelectric domains, and a triboelectric contribution from the encapsulated geometry combine to produce this enhancement.

## Experimental

### Synthesis of $HoFeO_3$ (HFO) Nanofibers

$HoFeO_3$ nanofibers were synthesized by a sol-gel-assisted electrospinning method. Poly(vinyl pyrrolidone) (PVP, average $M_w$ ~1,300,000; 1 g) was dissolved in 6 mL anhydrous ethanol under magnetic stirring for 2 h to form a homogeneous polymer solution. Separately, stoichiometric quantities of $Ho(NO_3)_3 \cdot 5H_2O$ and $Fe(NO_3)_3 \cdot 9H_2O$ were co-dissolved in a 1:1 (v/v) ethanol-water mixture (10 mL total) at a combined metal-ion concentration of 0.39 M (Ho:Fe molar ratio = 1:1) and stirred for 12 h to ensure complete dissolution and metal-ion homogenization. The metal-nitrate solution was then combined with the PVP solution and stirred for an additional 2 h to yield a viscous, sprayable sol. The sol was loaded into a stainless-steel syringe and electrospun under an applied voltage of 15 kV, a volumetric flow rate of 1 mL $h^{-1}$, a needle-to-collector distance of 11 cm, and a rotating-drum collector speed of 800 rpm to promote fiber alignment. As-spun composite fibers were dried at 120 °C for 12 h to remove residual solvent and then annealed in air at 800 °C and 850 °C for 3 h (ramp rate 5 °C $min^{-1}$) to pyrolyze the PVP template and crystallize the orthorhombic $HoFeO_3$ phase.

### Synthesis of $BaTiO_3$ (BTO) Nanofibers

$BaTiO_3$ nanofibers were prepared by a separate sol-gel electrospinning route. Barium acetate (1.275 g) was dissolved in 3 mL glacial acetic acid under stirring for 30 min. Titanium (IV) tetraisopropoxide (1.475 mL) was added dropwise to the barium acetate solution under continuous stirring to prevent premature hydrolysis; the Ba:Ti molar ratio was maintained at exactly 1:1. This solution was combined with a separately prepared PVP solution (0.2 g in 3 mL ethanol) and stirred for 2 h to yield a clear, viscous sol. Electrospinning was performed under identical conditions to HFO (15 kV, 1 mL $h^{-1}$, 11 cm needle-to-collector distance, 800 rpm rotating aluminum-foil collector). As-spun fibers were dried at 120 °C for 12 h and annealed in air at 750°C and 800°C for 3 h (ramp 5 °C $min^{-1}$) to crystallize the tetragonal $BaTiO_3$ phase.

### Synthesis of $HoFeO_3$-$BaTiO_3$ (HFO-BTO) Heterostructure Nanofibers

HFO-BTO heterostructure nanofibers were fabricated using an alternating electrospinning strategy designed to produce layered nanofiber mats with intimate HFO-BTO phase integration. The HFO and BTO precursor solutions were prepared as described above and loaded into separate syringes. The two solutions were electrospun alternately onto the same rotating collector in successive 1 h intervals (HFO, then BTO, then HFO) until the desired mat thickness was achieved. All other electrospinning parameters were identical to those used for the single-phase fibers. The collected multi-layered mat was dried at 120 °C for 12 h and annealed at 800 °C for 3 h in air to simultaneously crystallize both phases and burn out the PVP, yielding the final heterostructure nanofibers. This alternating approach ensures that individual HFO and BTO fibers are in proximity and share direct contact interfaces upon annealing, maximizing the interfacial area and the associated strain coupling. Related biphasic ceramic nanofibers, in which two oxide phases are integrated side by side within individual fibers, have been obtained by modified electrospinning routes. [24]

## Structural Characterization

Room-temperature powder X-ray diffraction (XRD) patterns of HFO, BTO, and HFO-BTO samples were recorded on a Bruker AXS D8 ADVANCE diffractometer operating in Bragg-Brentano geometry (Cu K$\alpha$ radiation, $\lambda = 1.5406$ Å) over a $2\theta$ range of 20°-80°. Lattice parameters and structural refinement were carried out using the Rietveld method as implemented in the FULLPROF suite, with crystal structure models for orthorhombic $HoFeO_3$ (*Pbnm*, No. 62) and tetragonal $BaTiO_3$ (*P4mm*, No. 99) as starting models. Transmission electron microscopy (TEM) and high-resolution TEM (HRTEM) were performed on a JEOL JEM-2100F instrument operating at 200 kV; samples were ultrasonically dispersed in ethanol and drop-cast onto holey carbon-coated copper grids. Raman spectra and mapping were acquired on a Horiba Jobin Yvon LabRAM instrument ($\lambda_{ex} = 532$ nm, power 5 mW) over a scan area of ~5 μm² using a 0.90/100× objective (spatial resolution 361 nm, depth resolution ~1 μm).

## Magnetic and Ferroelectric Characterization

DC magnetization measurements were performed on a Quantum Design Physical Property Measurement System (PPMS, Model 6000) using zero-field-cooled (ZFC) and field-cooled (FC) protocols at an applied field of 100 Oe over the temperature range 2-400 K. Full magnetization-versus-field (*M*-*H*) hysteresis loops were recorded at 5 K and 300 K. Ferroelectric polarization-electric-field (*P*-*E*) measurements were performed on an aix ACCT TF Analyzer 2000. Polarization and switching-current loops were recorded at room temperature at ±20 kV $cm^{-1}$ and 1.25 Hz for the BTO and HFO-BTO samples, and at ±10 kV $cm^{-1}$ and 1.5 Hz for the HFO sample. These measurements were performed on the calcined HFO, BTO, and HFO-BTO nanofiber powders, not on the PDMS composite films used in the devices.

## Device Fabrication and Performance Evaluation

Nanogenerator devices based on HFO, BTO, and HFO-BTO heterostructure nanofibers were fabricated using a PDMS encapsulation approach. A loading fraction of 5 wt% nanofibers or nanopowders was dispersed uniformly in uncured PDMS base resin under vigorous mechanical stirring for 30 min, followed by addition of the PDMS curing agent (10:1 base: agent by mass) and degassing under vacuum for 20 min to remove entrapped air bubbles. The mixture was solvent-cast into uniform films and thermally cured at 70 °C for 3 h. The cured composite films (~700 μm thick) were cut into 2 × 2 $cm^2$ squares and sandwiched between two indium-tin-oxide (ITO)-coated poly (ethylene terephthalate) (PET) electrodes, with silver paste contacts at the electrode edges. Devices were excited cyclically using an indigenously built sewing-machine-based mechanical press applying a force of ~40 N at a frequency of 14 Hz. Because the active composite film is sandwiched between two electrodes and operated in a contact-separation mode, the measured electrical signal contains both a piezoelectric component (from strain-induced polarization change in the $BaTiO_3$-rich, strain-tuned dipole regions) and a triboelectric component (from contact electrification at the nanofiber-PDMS and film-ITO electrode interfaces); the device therefore operates as a piezoelectric-dominated hybrid nanogenerator. [29,32] Open-circuit voltage and short-circuit current were recorded with a Tektronix MSO 2024 oscilloscope and a Keithley DMM 7510 digital multimeter, respectively. Power density was calculated as $P_V = V_L^2/(R_L V_d)$, where $V_L$ is the peak voltage across the load resistance $R_L$ and $V_d$ = 0.28 cm$^3$ is the active device volume (2 × 2 cm$^2$ × 700 µm).

## Results and Discussion

### X Ray Diffraction

The phase formation of the HFO, BTO, and HFO-BTO nanofibers was examined by XRD with Rietveld refinement (Figure 1). The BTO fibers index fully to tetragonal $BaTiO_3$ (*P4mm*, No. 99). The HFO fibers index to orthorhombic $HoFeO_3$ (*Pbnm*, No. 62), with trace $Ho_2O_3$ and α-$Fe_2O_3$ reflections attributed to incomplete orthoferrite crystallization at the annealing temperature used. The heterostructure pattern contains reflections of both HFO and BTO, confirming their coexistence in the nanofiber mat. No additional phases from reaction between the two oxides (e.g. Ba-Fe-O or Ho-Ti-O compounds) are detected. The pseudo-cubic lattice parameter of $HoFeO_3$ (~3.83 Å) is ~4% smaller than a parameter of $BaTiO_3$ (~3.99 Å). BTO in contact with HFO is therefore expected to experience in-plane compression, which in perovskite films favors the tetragonal distortion [11]. Consistent with this, the BTO reflections of the heterostructure show no resolvable shift (≤0.02°) and no measurable broadening (FWHM of the (110) reflection ~0.42° vs ~0.47° for pure BTO) relative to the single-phase reference.

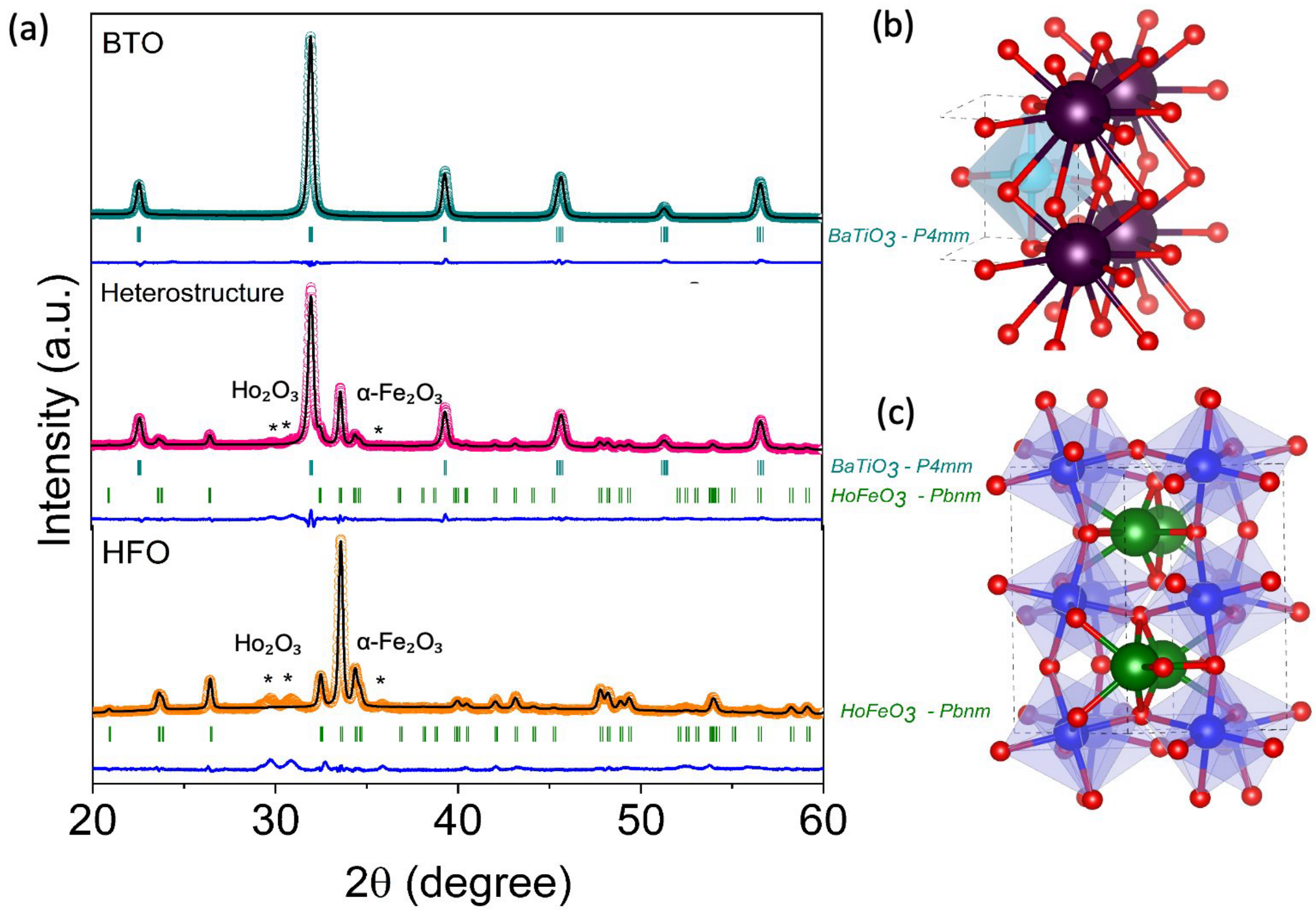

**Figure 1.** (a) Rietveld-refined XRD patterns of BTO, the HFO-BTO heterostructure, and HFO nanofibers (offset for clarity). Vertical tick marks indicate allowed reflections for tetragonal $BaTiO_3$ (*P4mm*) and orthorhombic $HoFeO_3$ (*Pbnm*); asterisks (*) denote trace $Ho_2O_3$ and α-$Fe_2O_3$ secondary-phase reflections. (b) Polyhedral crystal structure of tetragonal $BaTiO_3$ (*P4mm*), showing the off-center displacement of $Ti^{4+}$ within the $TiO_6$ octahedron. (c) Crystal structure of orthorhombic $HoFeO_3$ (*Pbnm*), with corner-sharing $FeO_6$ octahedra and $Ho^{3+}$ in the distorted A-site cavity (effectively eight-coordinate).

## Electron Density and Ti-O Covalency Enhancement

To rationalize the observed interface-driven enhancement of the ferroelectric response at the microscopic level, we compared the calculated valence charge-density distributions of the bulk HFO and BTO phases with those of the HFO/HFO-BTO and BTO/HFO-BTO interfacial slabs along the (100) and (010) crystallographic projections (Figure 2). This approach, which follows established methodology for strain-engineered perovskite films, [14,15] allows direct visualization of how orbital overlap and bonding character change at the heterointerface relative to the respective bulk references.

At the heterointerface, the valence charge density around the Ti and O sites is redistributed relative to bulk BTO, with increased density in the Ti-O bond region (Figure 2b3, d3, f3). Such bond-centered charge is characteristic of Ti 3*d*-O 2*p* hybridization beyond a purely ionic picture. [9,10] Because the ferroelectric instability of $BaTiO_3$ is driven by this Ti-O covalency, which softens the polar mode and stabilizes the off-center configuration, [9] a stronger interfacial Ti-O overlap is expected to increase the $Ti^{4+}$ displacement and the local dipole moment. On this basis, interfacial strain modifies not only the lattice parameters but also the electronic structure, in a way that favors polarization. This provides a plausible microscopic origin for the strain-tuned dipole enhancement. A comparable sensitivity of polarization to strain has been reported for strained perovskite films. [14,15]

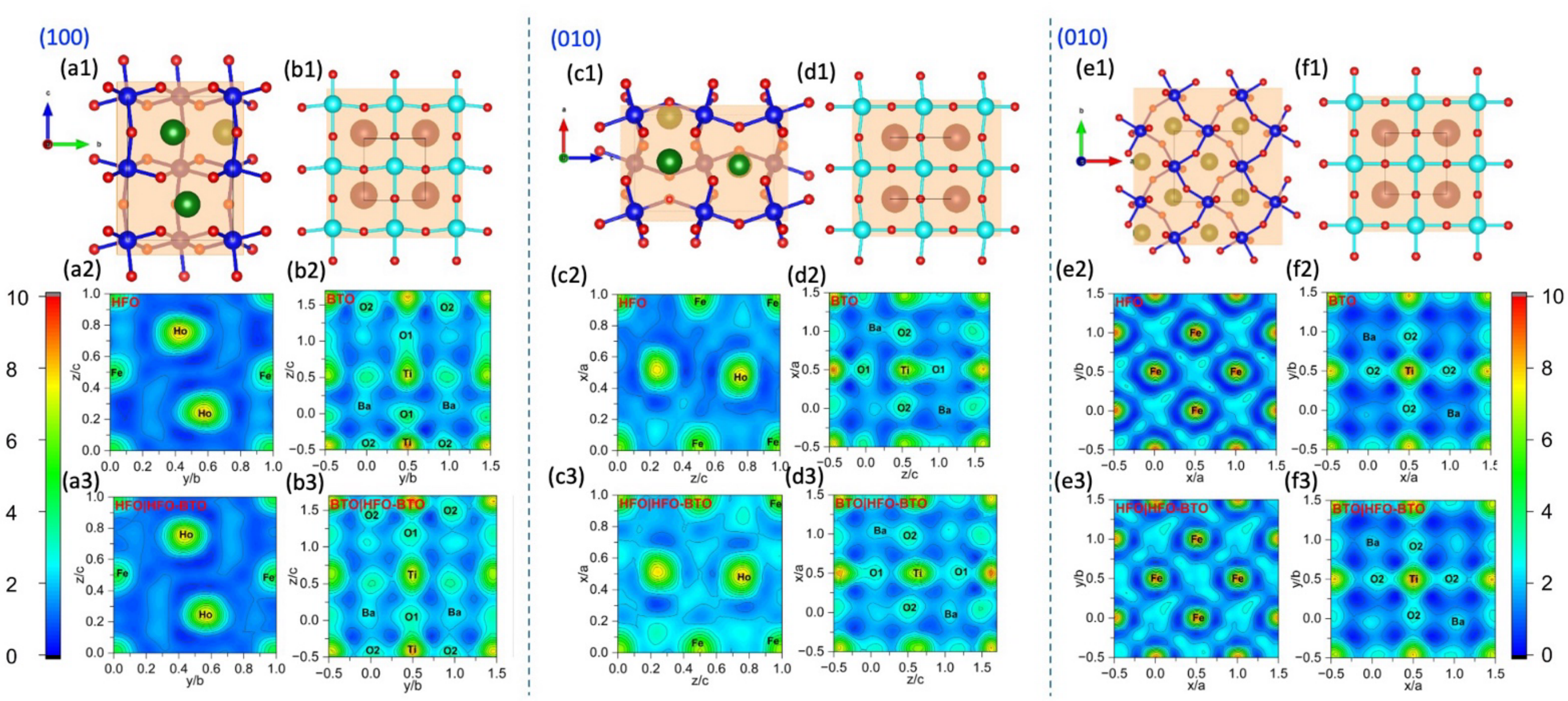

**Figure 2.** Calculated valence charge-density maps for (left to right) bulk HFO, the HFO|HFO-BTO interface, the BTO|HFO-BTO interface, and bulk BTO, viewed along the (100) and (010) projections. The pronounced accumulation of charge density in the Ti-O bond region at the heterointerface (compared with bulk BTO) indicates strengthened Ti 3$d$-O 2$p$ hybridization at the interface.

## Morphology, Microstructure, and Elemental Distribution

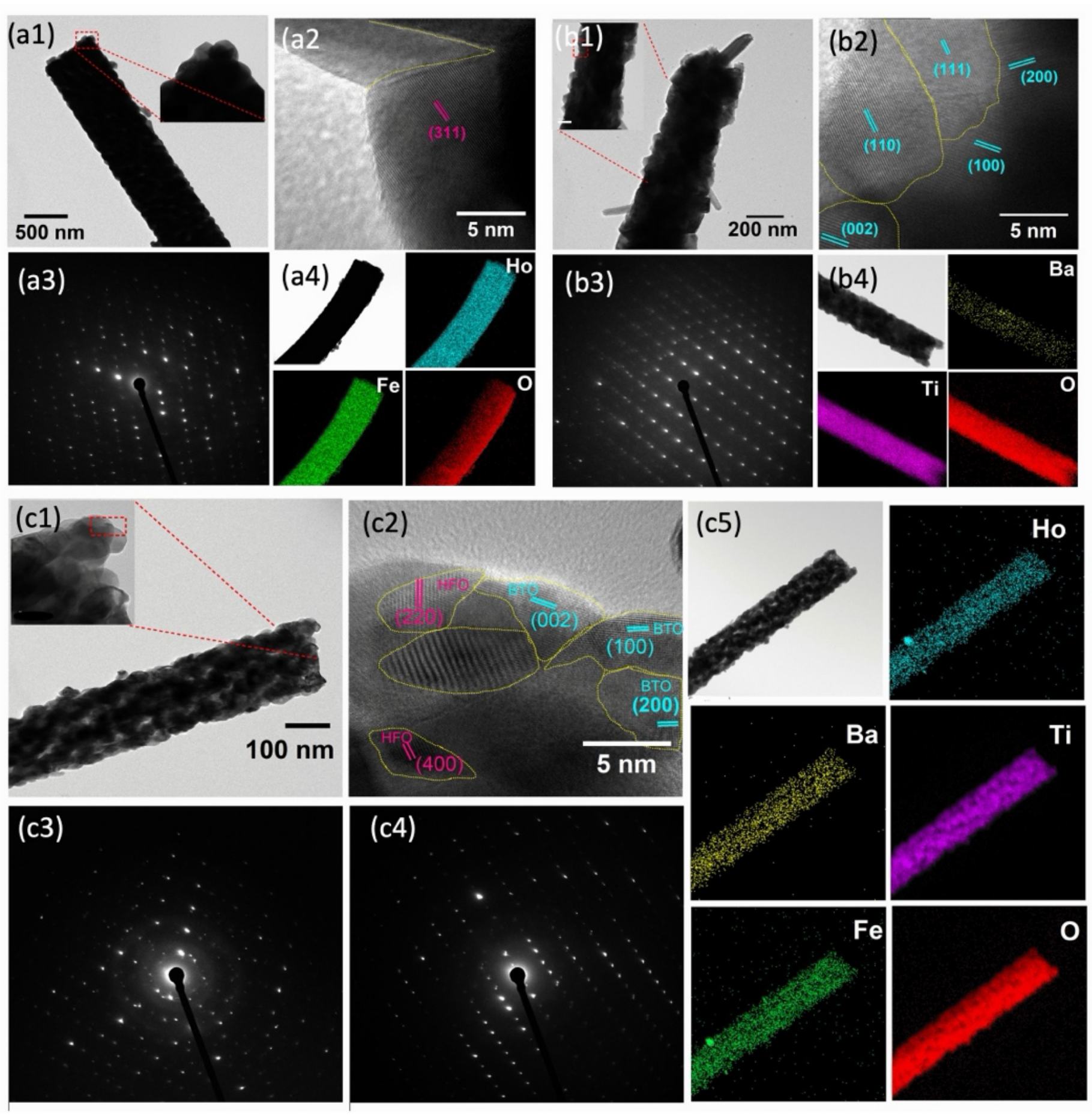

**Figure 3.** TEM, HRTEM, SAED, and EDS elemental mapping of (a) $HoFeO_3$, (b) $BaTiO_3$, and (c) HFO-BTO heterostructure nanofibers. HRTEM panels (a2, b2, c2) show lattice fringes indexed to the respective phases; (a3, b3, c3, c4) are SAED patterns confirming crystallinity and coherent dual-phase integration; (a4, b4, c5) are EDS elemental maps (Ho, Fe, O for HFO; Ba, Ti, O for BTO; all five elements for HFO-BTO) demonstrating uniform composition and successful heterostructure formation.

TEM images show continuous, high-aspect-ratio nanofibers for all three samples (Figure 3). SEM analysis of fibers give average diameters of ~670 nm (HFO), ~345 nm (BTO), and ~150 nm (HFO-BTO) (Figure S1). The fibers are polycrystalline, composed of grains of ~4-9 nm (crystalline domains in HRTEM). HRTEM lattice fringes of the HFO fibers index to orthorhombic $HoFeO_3$ (e.g. (311)), and those of the BTO fibers to tetragonal $BaTiO_3$ (e.g. (002), (100), (110), (111), (200)). In the HFO-BTO fibers, grains indexed to $HoFeO_3$ ((220), (400)) and $BaTiO_3$ ((002), (100), (200)) occur side by side within the same fiber and share grain boundaries (Figure 3c2). This provides the intimate HFO|BTO contact required for interfacial strain coupling. SAED patterns (Figure 3a3, b3, c3, c4) confirm the crystallinity of all samples. EDS maps show Ho, Fe, Ba, Ti, and O distributed along the heterostructure fibers, with a few Ho/Fe-rich spots, confirming dual-phase fibers without macroscopic phase segregation.

## Vibrational Signatures and Octahedral Distortion (Raman Spectroscopy)

Raman spectroscopy provides a sensitive, non-destructive probe of local bonding changes, octahedral tilting, and lattice distortions at a length scale shorter than the XRD coherence length and thus offers complementary evidence for the interfacial strain coupling inferred from the charge-density calculations (Figure 4). All vibrational modes observed in the HFO-BTO heterostructure can be assigned to either orthorhombic $HoFeO_3$ or tetragonal $BaTiO_3$ based on group-theoretical analysis and comparison with the literature (Table 1). No additional modes attributable to reaction products are observed. However, the strongest $\alpha$-$Fe_2O_3$ modes (~225, 245, 290, 410, and 610 $cm^{-1}$) overlap $HoFeO_3$ bands, so Raman cannot exclude the trace hematite detected by XRD.

Relative to the single-phase references, the heterostructure spectrum shows mode shifts of ~1-9 $cm^{-1}$ and substantial broadening (Figure 4, Table 1). The $A_1(TO_3)$ Ti-O stretching mode shifts from ~521 $cm^{-1}$ in BTO to ~524 $cm^{-1}$ in the heterostructure, consistent with the shorter, stiffer Ti-O bond suggested by the charge-density analysis (Figure 2). At the same time, the sharp $B_1$+E mode at ~305 $cm^{-1}$, a signature of long-range tetragonal order in $BaTiO_3$, broadens in the heterostructure and merges with neighboring bands, and the $A_1(TO_2)$ and $A_1(TO_3)$ bands broaden markedly. We attribute this to a distribution of local strain environments across the HFO|BTO interfaces, together with the smaller grain size. It indicates that the tetragonal distortion in the heterostructure is locally strained and less coherent over long range. The low-frequency $A_g$ mode of HFO likewise shifts by ~9 $cm^{-1}$, reflecting changes in the Fe-O-Fe bond angles. Together, these vibrational changes indicate strong interfacial coupling between the HFO and BTO grains.

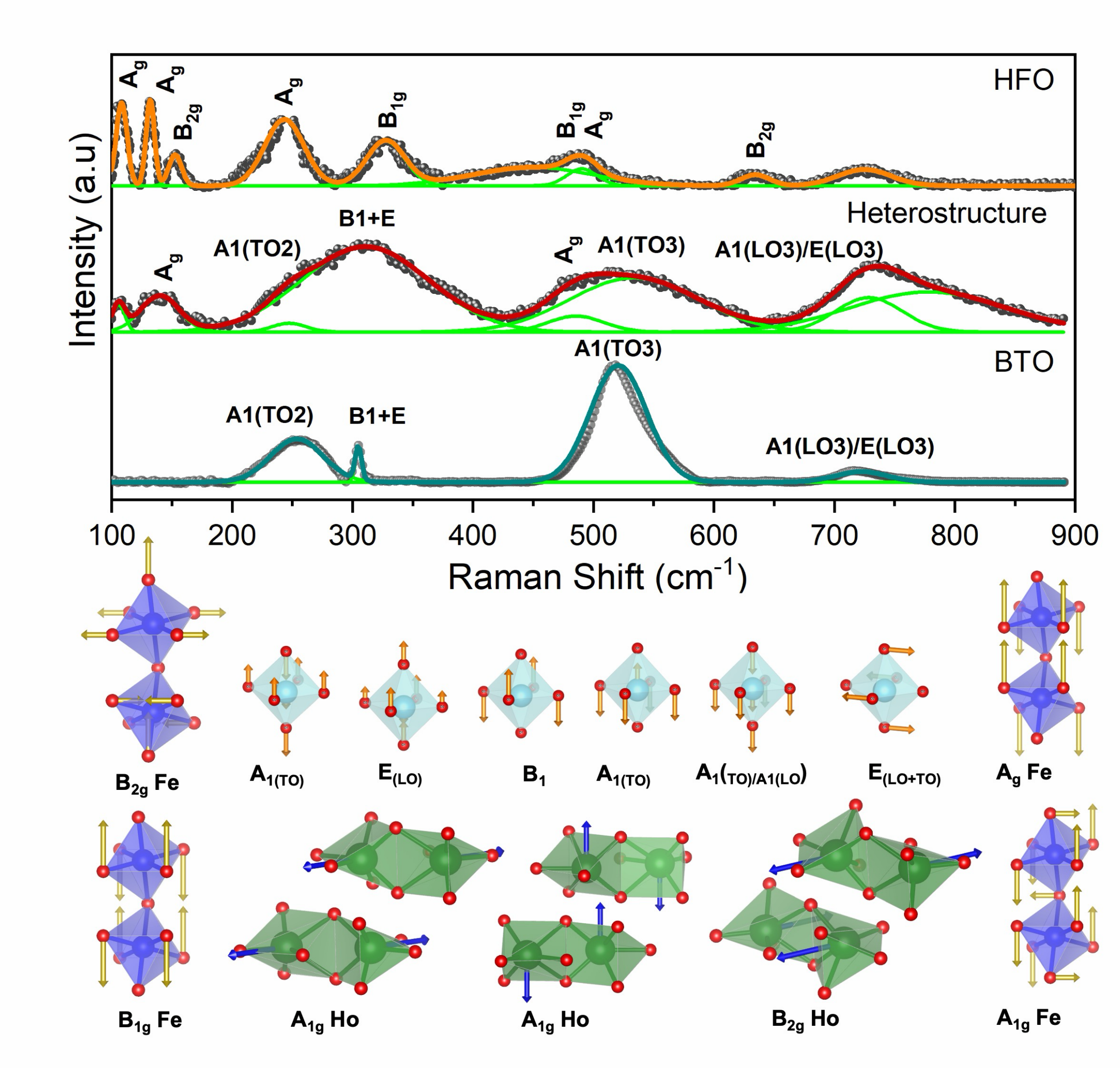

**Figure 4.** (Top) Deconvoluted Raman spectra of HFO, HFO-BTO heterostructure, and BTO nanofibers, with phase-wise mode assignment indicated. (Bottom) Schematic eigen-displacement patterns of the principal $FeO_6$ (left) and $TiO_6$ (right) vibrational modes, illustrating the atomic motions responsible for the observed peaks. Systematic shifts in the heterostructure spectra relative to single-phase references evidence interfacial strain-induced lattice distortion.

**Table 1.** Observed Raman modes with phase-wise assignment for HFO, BTO, and HFO-BTO nanofibers. Mode assignments follow refs [34] (orthoferrites) and [35] ($BaTiO_3$).

| Reported ($cm^{-1}$) | Sym. | Assignment | HFO NFs ($cm^{-1}$) | BTO NFs ($cm^{-1}$) | HFO-BTO NFs ($cm^{-1}$) | Ref. |
|---|---|---|---|---|---|---|
| ~110 | Ag | Ho-ion translational (R-mode) | 110 | | 112 | [34] |
| ~136 | Ag | Rare-earth cage motion | 133 | | 135 | [34] |
| ~158-160 | B2g | $FeO_6$ rotation | 155 | | 157 | [34] |
| ~223-225 | Ag | $FeO_6$ rotation | 223 | | 225 | [34] |
| ~278-290 | B1g | $FeO_6$ tilting/rotation | 284 | | 286 | [34] |
| ~346-350 | Ag | Fe-O stretching | 346 | | 348 | [34] |
| ~405-410 | | R-O vibration | 407 | | 409 | [34] |
| ~450-455 | B1g | $FeO_6$ bending | 475 | | 472 | [34] |
| ~515-520 | | $FeO_6$ deformation | 516 | | 514 | [34] |
| ~640-650 | B2g | $FeO_6$ stretching | 644 | | 646 | [34] |
| ~180 | E(TO) | Ba-Ti-O vibration | | 181 | 184 | [35] |
| ~260-270 | A1(TO) | Ti-O bending | | 268 | 271 | [35] |
| ~305 | E(TO) | Ti-O vibration | | 304 | 308 | [35] |
| ~515 | A1(TO) | Ti-O stretching | | 514 | 518 | [35] |
| ~715-720 | A1(LO) | Ti-O stretching (LO) | | 719 | 724 | [35] |

## Magnetic Behavior and Interfacial Exchange Coupling

The magnetic response of the HFO-BTO nanofibers arises predominantly from the $HoFeO_3$ phase (Figure 5). At 300 K it is governed by $Fe^{3+}$-O-$Fe^{3+}$ superexchange within the corner-sharing $FeO_6$ network. [16,17] In the Goodenough-Kanamori-Anderson framework, this superexchange depends sensitively on the Fe-O-Fe bond angles and Fe-O bond lengths. The octahedral tilting of the *Pbnm* structure produces a canted antiferromagnetic arrangement with a weak ferromagnetic moment along the *c* axis. [16] The HFO loop at 300 K shows this weak ferromagnetism superimposed on a linear antiferromagnetic contribution. Trace α-$Fe_2O_3$, itself weakly ferromagnetic above its Morin transition (~260 K), may contribute a small part of the hysteresis. At 5 K the magnetization is dominated by the large paramagnetic moment of $Ho^{3+}$, polarized by the Fe sublattice, which accounts for the much larger, near-saturating magnetization. $HoFeO_3$ additionally undergoes a spin-reorientation transition at ~35-60 K, and nanoscale $RFeO_3$ shows size- and chemical-pressure-dependent magnetism. [18,19] The BTO fibers show a small, saturating signal (~0.1 emu $cm^{-3}$ at 300 K), typical of defect-related (e.g. oxygen-vacancy) magnetism in nanostructured $BaTiO_3$ [36]; this is negligible compared with HFO. The high-field magnetization of the heterostructure is ~23% of that of pure HFO at both 300 K and 5 K, consistent with the HFO fraction of the material. Its coercive field is similar to that of HFO at 300 K (~0.11 T vs ~0.10 T) but lower at 5 K (~0.17 T vs ~0.22 T), with a slightly asymmetric 5 K loop, indicating that interfacial strain modifies the $FeO_6$ distortion and spin canting while preserving the overall antiferromagnetic order.

The retention of the orthoferrite weak-ferromagnetic hysteresis in the heterostructure confirms that the $HoFeO_3$ $FeO_6$ framework survives electrospinning and co-annealing with BTO without significant reaction. Although magnetic order does not contribute to the nanogenerator output, the chemical and structural integrity of the HFO phase is a prerequisite for the interfacial strain coupling discussed above.

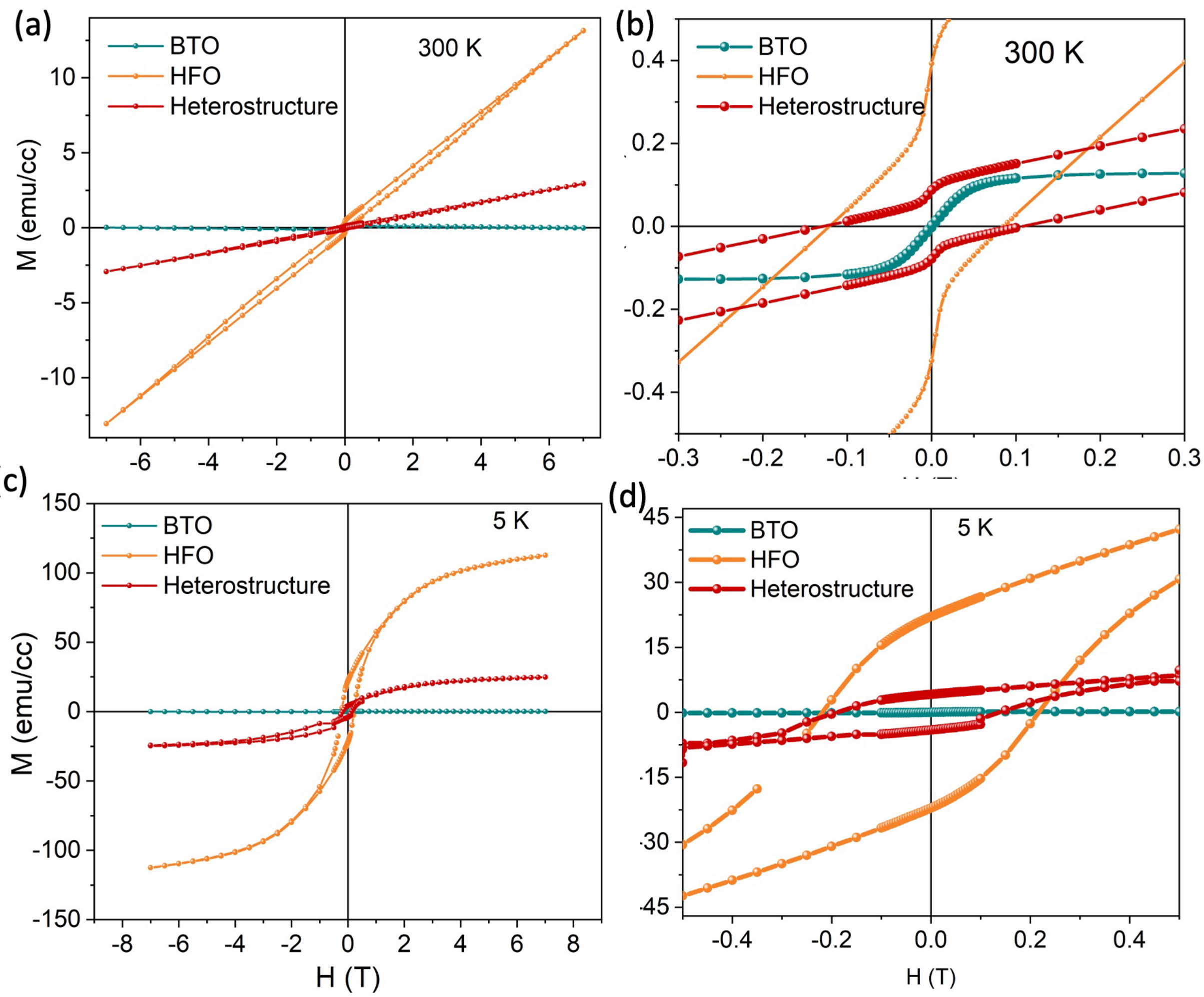

**Figure 5.** Magnetization-field (*M-H*) loops of BTO, HFO, and HFO-BTO nanofibers at (a) 300 K and (c) 5 K, with magnified low-field views (b) and (d) highlighting the weak ferromagnetic hysteresis. The heterostructure retains a robust orthoferrite-derived hysteresis loop, confirming structural integrity of the $FeO_6$ framework after co-annealing.

### Ferroelectric Polarization and Switching Current

Polarization–electric field ($P$–$E$) loops and the corresponding switching-current density ($J$–$E$) loops of the three samples are shown in Figure 6. The BTO sample shows a polarization of 15.80 μC $cm^{-2}$ at zero field, with current maxima of ~±2.3 μA $cm^{-2}$ at ~±3.5 kV $cm^{-1}$, consistent with domain switching in the tetragonal $BaTiO_3$ grains. The HFO-BTO heterostructure shows a comparable zero-field polarization of 16.39 μC $cm^{-2}$, but its switching current is broader, lower, and shifted to a higher field (~±1.7 μA $cm^{-2}$ at ~±7 kV $cm^{-1}$), i.e. the apparent coercive field roughly doubles. A broader switching response at higher field is consistent with domains that are partly clamped by the interfacial compressive strain inferred from the Raman data (Figure 4). Such strain stabilizes the tetragonal distortion and makes the polarization harder to reverse. For piezoelectric harvesting this is advantageous, because the output comes from strain-induced changes of an already oriented polarization rather than from switching, and a harder polarization resists back-switching and depolarization under repeated loading. Because the loops were measured on the nanofiber powders rather than on the 5 wt% PDMS composites, they characterize the switching behavior of the fibers themselves; inside the devices, the fibers experience a much smaller local field and are not switched during harvesting.

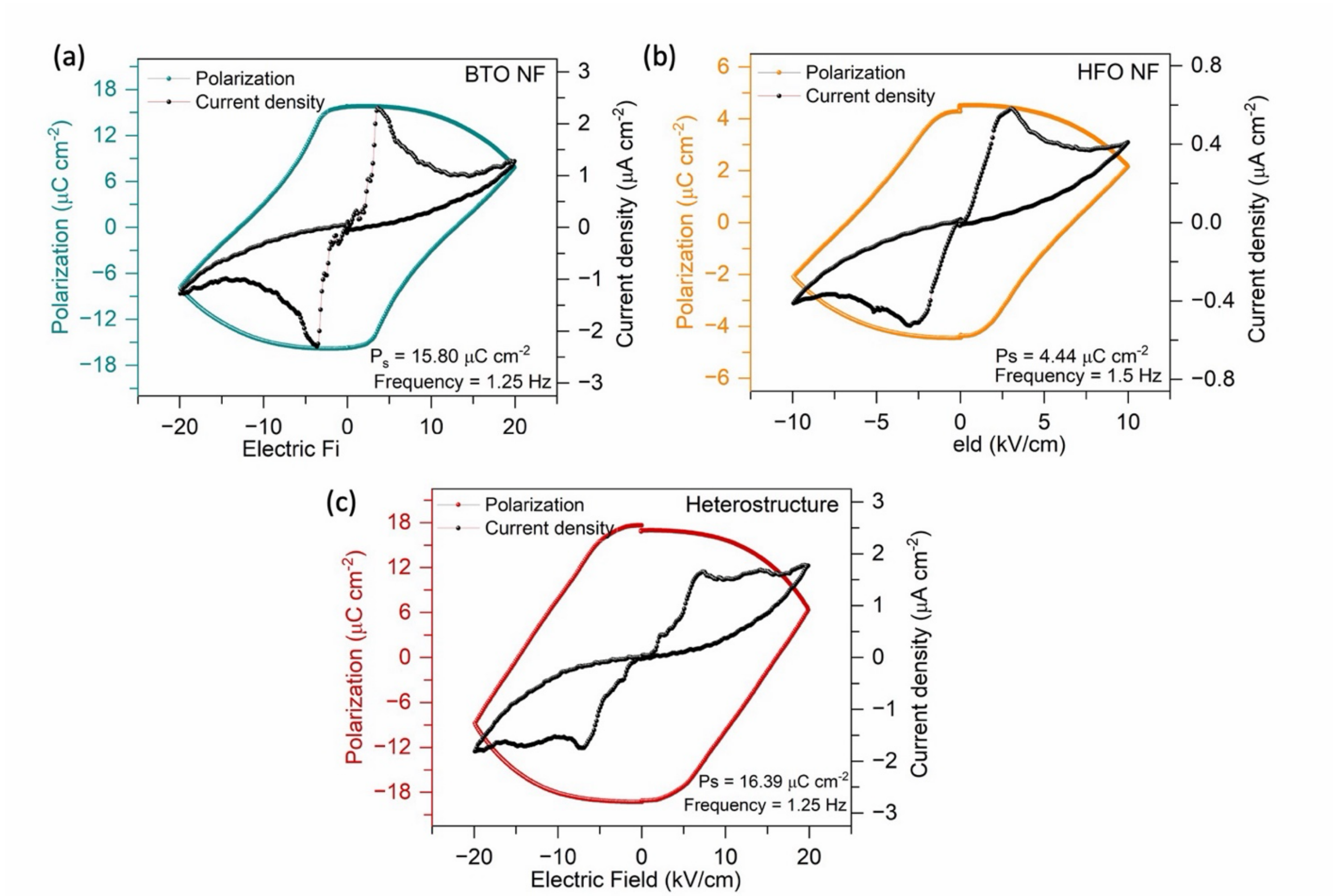

**Figure 6.** Polarization–electric field ($P$–$E$, colored) and switching-current density ($J$–$E$, black) loops of (a) BTO, (b) HFO, and (c) HFO-BTO heterostructure nanofiber powders, recorded at room temperature at ±20 kV $cm^{-1}$ and 1.25 Hz for BTO and HFO-BTO, and at ±10 kV $cm^{-1}$ and 1.5 Hz for HFO. The values in each panel (15.80, 4.44, and 16.39 μC $cm^{-2}$) are the polarizations at zero field; they include a leakage contribution (see text).

Leakage through grain-boundary and defect-mediated conduction paths, charge trapping at oxygen vacancies and interface states, and depolarization fields at free surfaces all let piezoelectrically generated bound charge be screened internally rather than driven through the external circuit. [12,13] Leakage also shapes the present loops. All three are rounded, with the polarization largest near zero field and lower at the field maximum (~7–8 μC $cm^{-2}$ for BTO and ~6–7 μC $cm^{-2}$ for HFO-BTO at +20 kV $cm^{-1}$), and the fields at which the polarization crosses zero (~±13 kV $cm^{-1}$ for BTO, ~±15–16 kV $cm^{-1}$ for HFO-BTO) lie well above the switching-current maxima, whereas in an ideal ferroelectric the two coincide. [37,38] The HFO sample confirms a substantial non-ferroelectric contribution: $HoFeO_3$ is centrosymmetric at room temperature, yet it shows an apparent zero-field polarization of 4.44 μC $cm^{-2}$ and current maxima of ~0.5 μA $cm^{-2}$ at ~±3 kV $cm^{-1}$. [39] Ferroelectric-like loops and current maxima of this kind are well documented in non-ferroelectric and leaky dielectrics, where they arise from conduction and from space-charge (Maxwell–Wagner) polarization at grain boundaries and electrode interfaces rather than from domain reversal. [39-41] Strong leakage is expected in the present samples for three reasons: (i) they are compacted powders with a high density of particle and grain boundaries, which favors interfacial polarization and boundary conduction; (ii) oxygen vacancies formed during calcination in air act as mobile charge carriers in nanocrystalline $BaTiO_3$; and (iii) in the orthoferrite, oxygen vacancies and mixed $Fe^{2+}/Fe^{3+}$ valence, together with the trace α-$Fe_2O_3$ detected by XRD, provide hopping conduction paths. [18] The zero-field polarizations should therefore be read as upper bounds that include conduction and interfacial (Maxwell–Wagner) charge, and the 4% difference between BTO and HFO-BTO lies within this uncertainty. The high-field current density is also somewhat larger for the heterostructure (~1.8 μA $cm^{-2}$ at ±20 kV $cm^{-1}$, compared with ~1.2–1.3 μA $cm^{-2}$ for BTO), so the interfaces do not reduce leakage under high bias. Leakage-compensated or positive-up-negative-down (PUND) measurements would be needed to isolate the true switching charge. [37,38] Within these limits, the *P–E* data show that the heterostructure retains a switchable polarization comparable to that of BTO with a higher coercive field, rather than a larger polarization. Its higher power output (Figure 7b, c) is therefore attributed mainly to the strain-modified $TiO_6$ environment and to more stable, less easily depolarized domains, and not to reduced leakage.

## Nanogenerator Performance and the Triboelectric Contribution

Nanogenerator performance under cyclic impact (~40 N, 14 Hz) is summarized in Figure 7b, c. The HFO-BTO device produces open-circuit voltage peaks of about +6 V and −8 V, compared with about +4/−5 V for BTO and ±2.6 V for HFO (Figure 7b). At a load resistance of 4.7 MΩ, the heterostructure device delivers a power density of 24.31 μW $cm^{-3}$ (Figure 7c). This is 2.4 times the BTO value (10.00 μW $cm^{-3}$) and 8.7 times the HFO value (2.78 μW $cm^{-3}$) and corresponds to power density of ~1.7 μW $cm^{-2}$. The corresponding current through the 4.7 MΩ load is ~1.20 μA (HFO-BTO), ~0.77 μA (BTO), and ~0.41 μA (HFO).

The primary energy-conversion pathway is expected to be piezoelectric. Cyclic stress strains the $BaTiO_3$ grains and changes their polarization, and the associated bound charge is compensated by free electrons flowing through the external circuit. Superimposed on this is a triboelectric contribution. During each contact and separation event, the difference in electron affinity between the nanofiber/PDMS composite surface and the ITO-coated PET counter-electrode drives contact-electrification charge transfer, which induces an electrostatic-induction current during separation. [29,31,32] The HFO device provides a useful internal reference for separating the two channels. $HoFeO_3$ is centrosymmetric and hence non-piezoelectric at room temperature, so its output under identical conditions (2.78 μW $cm^{-3}$) must arise from triboelectric, electret, and possibly flexoelectric effects. Because all devices share the same 5 wt% loading, PDMS matrix, and electrode configuration, this value approximates the non-piezoelectric contribution, which is only ~11% of the heterostructure power.

The oxide nanofibers can also modulate the triboelectric channel. A high-permittivity filler raises the effective permittivity of the composite through Maxwell-Wagner-Sillars interfacial polarization at the fiber-polymer boundaries, [28,33,42,43] increasing the charge that can be stored between contact events. The nanofiber morphology also increases the effective contact area and surface roughness. [23,44] Interfacial strain in the heterostructure may further alter surface states and trap depths relative to single-phase BTO. Overall, the device is best described as a hybrid nanogenerator in which the piezoelectric channel dominates, and the triboelectric channel adds a complementary contribution.

For the triboelectric component, the open-circuit voltage in contact-separation mode scales as $V_{OC} \sim \sigma x(t)/\varepsilon_0$, where σ is the triboelectric surface charge density and $x(t)$ the separation distance. [45] The power delivered to a load, $P = V_L^2/R_L$, is maximized when $R_L$ matches the internal impedance of the generator. For a capacitive source such as this ~700 μm, 2 × 2 $cm^2$ film ($C \sim 15$ pF for an effective $\varepsilon_r \sim 3$), the matched load depends on the duration of each impact pulse rather than on the 14 Hz repetition rate, so it must be determined experimentally from a load sweep. The voltage output is uniform from cycle to cycle over the recorded interval (Figure 7b). Table 2 compares the device with representative $BaTiO_3$-based nanogenerators.

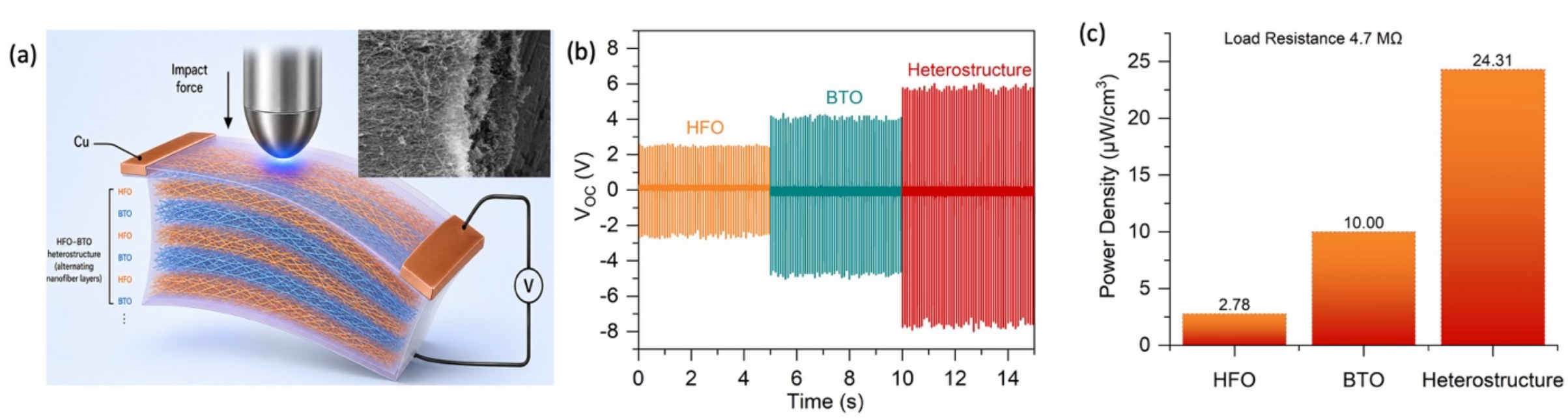

**Figure 7.** (a) Schematic cross-section of the flexible HFO-BTO hybrid PENG device, showing alternating HFO/BTO nanofiber layers embedded in PDMS between ITO-coated PET electrodes (inset: cross-sectional SEM of the nanofiber mat). (b) Open-circuit voltage ($V_{OC}$) versus time under cyclic mechanical impact (~40 N, 14 Hz) for HFO, BTO, and HFO-BTO devices, illustrating the superior and stable output of the heterostructure. (c) Power density at a 4.7 MΩ load: 24.31 μW $cm^{-3}$ (HFO-BTO), compared with 10.00 μW $cm^{-3}$ (BTO) and 2.78 μW $cm^{-3}$ (HFO).

**Table 2.** Comparison of representative $BaTiO_3$-based nanofiber and composite piezoelectric/hybrid nanogenerators from the literature with the present HFO-BTO device.

| **Active material** | **Mode** | **VOC** | **Power density** | **Ref.** |
|---|---|---|---|---|
| Aligned $BaTiO_3$ nanofibers + PDMS | Vertical compression | 2.67 V | 0.18 μW | [23] |
| $BaTiO_3$/PVDF electrospun nanofibers | Vertical compression | 50 V | 407 μW $cm^{-2}$ | [44] |
| $BaTiO_3$/PDMS composite (piezo-tribo hybrid) | Hybrid compression | 13.5 V | | [29] |
| High-ε $BaTiO_3$ NPs in sponge PDMS | Contact-separation | | | [28] |
| $BaTiO_3$/PDMS micropyramid (tribo-piezo) | Contact-separation | | High | [31] |
| **HFO-BTO nanofiber heterostructure + PDMS (this work)** | **Cyclic impact** | 14 V | 24.31 μW $cm^{-3}$ (~1.7 μW $cm^{-2}$) | **This work** |

### Structure-Morphology-Anisotropy-Performance Correlation

Most reported nanogenerators in the literature rely on polymeric piezoelectric matrices (e.g. PVDF and its copolymers) or unstructured particulate ceramics, where performance improvements are achieved primarily through surface functionalization, micro/nano-patterning of the polymer matrix, or compositional tuning of the ceramic filler. [28,31] The present HFO-BTO nanofiber heterostructure introduces a qualitatively different design paradigm: *interface-engineered oxide nanofibers with intrinsic anisotropy* that simultaneously optimize piezoelectric charge generation, charge retention, and long-term mechanical robustness. This approach can be understood through three linked design levels:

At the structural level, the coexistence of orthorhombic $HoFeO_3$ and tetragonal $BaTiO_3$ (Figure 1) produces interfacial lattice distortion. According to the calculations, this strengthens Ti-O covalency (Figure 2), and it is accompanied by a polarization comparable to that of BTO with a higher coercive field (Figure 6). At the morphological level, the high-aspect-ratio fibers (Figure 3) provide a large interfacial area with the PDMS matrix and efficient load transfer along the fiber length. At the materials level, the chemically robust orthoferrite phase survives co-annealing intact (Figures 1 and 5), preserving the HFO|BTO interfaces, while the $BaTiO_3$-rich regions dominate charge generation. [If fiber orientation within the PDMS film is controlled, add the anisotropy argument here with supporting SEM/orientation data.] This combination distinguishes the present all-oxide nanofiber heterostructure from single-phase ceramic fillers and polymer-piezoelectric composites.

### Mechanism of Nanogenerator Enhancement

The enhancement can be summarized as cooperative effects acting across length scales. At the atomic scale, interfacial strain strengthens Ti 3*d*-O 2*p* hybridization, favoring a larger off-center $Ti^{4+}$ displacement and local polarization (Figures 2 and 4). At the nanometer scale, interfacial strain partly clamps the ferroelectric domains, raising the coercive field while preserving a polarization comparable to that of BTO (Figure 6), which stabilizes the oriented polarization against depolarization under cyclic loading. At the micrometer scale, the high-aspect-ratio fibers transfer stress from the PDMS matrix to the oxide grains. At the device scale, the oxide-in-PDMS composite supports a secondary triboelectric contribution, estimated from the HFO reference device at ~11% of the total power. Together, these effects give the HFO-BTO device a power density of 24.31 $\mu W\ cm^{-3}$, 2.4 times that of the BTO control.

## Conclusions

We have prepared $HoFeO_3$-$BaTiO_3$ nanofiber heterostructures by alternating sol-gel electrospinning and demonstrated their use as the active phase of PDMS-encapsulated hybrid nanogenerators. The coexistence of orthorhombic $HoFeO_3$ and tetragonal $BaTiO_3$ within the fibers introduces interfacial lattice distortion. First-principles calculations indicate that this distortion strengthens Ti 3*d*-O 2*p* covalency at the interface, a mechanism that favors off-center $Ti^{4+}$ displacement. The heterostructure retains a polarization comparable to that of $BaTiO_3$ (16.39 vs 15.80 $\mu C\ cm^{-2}$ at zero field) with a higher coercive field, consistent with strain-stabilized ferroelectric domains that resist depolarization. Under cyclic impact, the device delivers open-circuit voltage peaks of about 14 V and a power density of 24.31 $\mu W\ cm^{-3}$, 2.4 times that of an identically prepared $BaTiO_3$ device. The triboelectric contribution from contact electrification is estimated at ~11% of the output. These results establish a structure-morphology-performance relationship for oxide nanofiber heterostructures. They show that interface-driven lattice engineering is an effective strategy for improving lead-free piezoelectric energy harvesters, with potential applications in self-powered sensing, wearable electronics, and structural health monitoring. [46]

## Author Contributions

## Conflicts of Interest

There are no conflicts to declare.

## Data Availability

All data supporting the conclusions of this article are disclosed within the manuscript and its Supplementary Information.

## Acknowledgements

The authors acknowledge the Indian Institute of Science Education and Research, Pune, for providing facilities for synthesis and characterization. S.C. acknowledges the Department of Science and Technology, India (grant DST TAR/2022/000621). R.B. acknowledges ANRF India (grant ANRF/ARG/2025/001573/CS).

## Notes and References